\documentclass[a4paper]{article}

\usepackage{amsmath}
\usepackage{amssymb}
\usepackage[margin=1.5cm,
            includefoot,
            footskip=30pt]{geometry}
\usepackage{layout}
\usepackage{graphicx}
\usepackage{subcaption}

\usepackage{biblatex}
\usepackage{pdfpages}
\usepackage{booktabs}

\usepackage{lscape}

\usepackage{authblk}

\bibliography{main.bib}
\newcommand{\SSe}{\text{SSE}}

\title{Recognising and evaluating the effectiveness
       of extortion in the Iterated Prisoner's Dilemma}
\author[1]{Vincent A. Knight*}
\author[2]{Marc Harper}
\author[1]{Nikoleta E. Glynatsi}
\author[1]{Jonathan Gillard}
\affil[1]{Cardiff University, School of Mathematics, Cardiff, United Kingdom}
\affil[2]{Google Inc., Mountain View, CA, United States of America}
\date{\today}

\begin{document}

\maketitle

\begin{abstract}

Since the introduction of zero-determinant strategies, extortionate strategies
have received considerable interest. While an interesting class of strategies,
the definitions of extortionate strategies are algebraically rigid, apply only
to memory-one strategies, and require complete knowledge of a strategy (memory-one
cooperation probabilities). We describe a method to detect extortionate
behaviour from the history of play of a strategy. When applied to a corpus of
204strategies this method
detects extortionate behaviour in well-known extortionate strategies as well
others that do not fit the algebraic definition. The highest performing
strategies in this corpus are able to exhibit selectively extortionate behavior,
cooperating with strong strategies while exploiting weaker strategies, which no
memory-one strategy can do. These strategies emerged from an evolutionary
selection process and their existence contradicts widely-repeated folklore in
the evolutionary game theory literature: complex strategies can be
extraordinarily effective, zero-determinant strategies can be outperformed by
non-zero determinant strategies, and longer memory strategies are able to
outperform short memory strategies. Moreover, while resistance to extortion is
critical for the evolution of cooperation, the extortion of weak opponents
need not prevent cooperation between stronger opponents, and this adaptability
may be crucial to maintaining cooperation in the long run.
\end{abstract}

The Iterated Prisoner's Dilemma is a model for rational and evolutionary
interactive behaviour, having applications in biology, the study of human
social behaviour, and many other domains. Since the introduction of
zero-determinant (ZD) strategies in \cite{Press2012}, extortionate strategies have
received considerable interest in the literature \cite{hilbe2015partners}.
These strategies ``enforce'' a difference in stationary
payouts between themselves and their opponents. The definition requires a
precise algebraic relationship between the probabilities of cooperation given
the outcome of the previous round of play and slight alterations to these
probabilities can cause a strategy to no longer satisfy the necessary equations.

In~\cite{adami2013evolutionary, Hilbe2013, hilbe2013adaptive, hilbe2015partners,
ichinose2018zero, Moran1707} the true effectiveness of these strategies in an
evolutionary setting was discussed. For example~\cite{adami2013evolutionary}
showed that ZD strategies were not evolutionarily stable. Furthermore, in that
work it was also postulated that `evolutionarily successful ZD strategies could
be designed that use longer memory to distinguish self from non-self'. In a non
evolutionary context, the work of~\cite{becks2019extortion} uses social
experiments to suggest that higher rewards promote extortionate
behaviour where statistical techniques are used to identify such behaviour.

The algebraic relationships of extortion define a subspace of
\(p\in\mathbb{R}^4\) which can be used broaden the definition of an extortionate
strategy by requiring only that the defining cooperation probabilities of a
strategy are close to an algebraically extortionate strategy, by the usual
technique of orthogonal projection. Moreover, given the history of play of a
strategy in an actual matchup, we can empirically observe its four
cooperation probabilities, measure the distance to the subspace of extortionate
strategies, and use this distance as a measure of the extortionality of a
strategy. This method can be applied to any strategy regardless of the memory
depth and avoids the algebraic rigidity issues.

We apply this method to the largest known corpus of strategies for the iterated
prisoner's dilemma (the Axelrod Python library~\cite{Knight2016, Knight2018})
and show empirically that the method in fact detects extortionate strategies.
A large tournament with 204
strategies demonstrates that sophisticated
strategies can in fact recognise extortionate behaviour and adapt to their
opponents. Further, statistical analysis of these strategies in the context of
evolutionary dynamics demonstrates the importance of adaptability to achieve
evolutionary stability. All of the code and data discussed in
Section~\ref{sec:numerical-experiments} is open sourced, archived, and written
according to best scientific principles~\cite{Wilson2014}. The data archive can
be found at~\cite{vincent_knight_2018_1297075} and the source code was developed
at~\url{https://github.com/drvinceknight/testing_for_ZD/} and has been archived
at~\cite{vincent_knight_2019_2598534}. In
Section~\ref{sec:evolutionary-dynamics}, this large tournament is complemented
with evolutionary dynamics that offer some insight in to the
effectiveness of extortionate strategies.

Several theoretical insights emerge from this work. Infamously, extortionate
strategies do not play well with themselves. In \cite{Press2012},
Press and Dyson claim that a player with a ``theory of mind'' would
rationally chose to cooperate against an opponent that also has knowledge
of zero-determinant strategies to avoid sustained mutual defection. While not
possible for memory-one strategies, we show that this behavior is exhibited by
relatively simple longer memory strategies which previously emerged from an
evolutionary selection process. Similarly, in
\cite{adami2013evolutionary}, Adami and Hintze suggest that there may exist
strategies that are able to selectively behave extortionately to some opponents
and cooperatively to others. We show that this is indeed the case for the same
evolved strategies. It seems that humans have trouble explicitly creating such
strategies but evolution is able to simply by optimizing for total payoff in IPD
interactions. Accordingly, while resistance to extortionate behavior appears
critical to the evolution of cooperation, there is no prohibition on selectively
extorting weaker opponents, even in population dynamics, and this behavior is
evolutionarily advantageous.

\section{Methods: Recognising Extortion}\label{sec:sserror-zd-strategies}

Zero-determinant strategies are a special case of memory-one strategies,
which are defined by elements of \(\mathbb{R}^4\) mapping a state of
\({\{C, D\}}^2\), corresponding to the prior round of play, to a probability of
cooperating in the next round. A match between two such strategies creates a
Markov chain with transient states \({\{C, D\}}^2\). The main result
of~\cite{Press2012} is that given two memory-one players \(p,
q\in\mathbb{R}^4\), a linear relationship between the players' scores can, in
some cases, be forced by one of the players for specific choices of these
probabilities.

Using the notation of~\cite{Press2012}, the utilities for player \(p\)
are given by \(S_x=(R, S, T, P)\) and for player \(q\) by \(S_y=(R, T, S, P)\)
and the stationary scores of each player are given by \(S_X\) and \(S_Y\)
respectively. The main result of~\cite{Press2012} is that if

\begin{equation}\label{eqn:linear_relationship_for_p}
    \tilde p=\alpha S_x + \beta S_y + \gamma
\end{equation}

or

\begin{equation}\label{eqn:linear_relationship_for_q}
    \tilde q=\alpha S_x + \beta S_y + \gamma
\end{equation}

 where \(\tilde p = (1 - p_1, 1 - p_2, p_3, p_4)\) and
\(\tilde q = (1 - q_1, 1 - q_2, q_3, q_4)\) then:

\begin{equation}
    \alpha S_X + \beta S_Y + \gamma = 0
\end{equation}

Extortionate strategies are defined as follows. If this relationship is
satisfied

\begin{equation}\label{eqn:constraint_for_extortion}
    \gamma = - P(\alpha + \beta)
\end{equation}

then the player can ensure \((S_X - P)=\chi(S_Y-P)\) where:

\begin{equation}\label{eqn:definition_of_chi}
    \chi=\frac{-\beta}{\alpha}
\end{equation}

\noindent Thus, if (\ref{eqn:constraint_for_extortion}) holds and \(\chi >1\) a player is
said to extort their opponent.
First, the reverse problem is considered: given a
\(p\in\mathbb{R}^4\) can one determine if the associated strategy is attempting
to act in an extortionate way?

\subsection{Subspace of Extortionate Strategies}

Constraints (\ref{eqn:linear_relationship_for_p}) and
(\ref{eqn:constraint_for_extortion}) correspond to:

\begin{align}
    \tilde p_1 & = \alpha R + \beta R - P (\alpha + \beta)
            \label{eqn:condition_for_tilde_p1}\\
    \tilde p_2 & = \alpha S + \beta T - P (\alpha + \beta)
            \label{eqn:condition_for_tilde_p2}\\
    \tilde p_3 & = \alpha T + \beta S - P (\alpha + \beta)
            \label{eqn:condition_for_tilde_p3}\\
    \tilde p_4 & = \alpha P + \beta P - P (\alpha + \beta) = 0
            \label{eqn:condition_for_tilde_p4}
\end{align}

Equation (\ref{eqn:condition_for_tilde_p4}) ensures that \(p_4=\tilde p_4=0\).
Equations (\ref{eqn:condition_for_tilde_p1}-\ref{eqn:condition_for_tilde_p3})
can be used to eliminate \(\alpha, \beta\), giving:

\begin{equation}\label{eqn:planar_definition_of_extortion}
    \tilde p_1 = \frac{(R - P)(\tilde p_2 + \tilde p_3)}{S + T - 2P}
\end{equation}

with:

\begin{equation}\label{eqn:definition_of_chi}
    \chi = \frac{\tilde p_2 (P - T) + \tilde p_3 (S - P)}
                {\tilde p_2 (P - S) + \tilde p_3 (T - P)}
\end{equation}

Given a strategy \(p\in\mathbb{R}^{4}\) equations
(\ref{eqn:condition_for_tilde_p4}-\ref{eqn:definition_of_chi}) can be used to
check if a strategy is extortionate. The conditions correspond to:

\begin{align}
    p_1 & = \frac{(R-P)(p_2 + p_3) - R + T + S - P}{S + T - 2P}
     \label{eqn:condition_for_p1}\\
    p_4 & = 0 \label{eqn:condition_for_p4}\\
    1 & > p_2 + p_3\label{eqn:condition_for_chi}
\end{align}

The algebraic steps necessary to prove these results are available in the
supporting materials, and note that an equivalent formulation was obtained
in~\cite{adami2013evolutionary}.

All extortionate strategies reside on a triangular (\ref{eqn:condition_for_chi})
plane (\ref{eqn:condition_for_p1}) in 3 dimensions (\ref{eqn:condition_for_p4}).
Using this formulation it can be seen that a necessary (but not sufficient)
condition for an extortionate strategy is that it cooperates on average less
than 50\% of the time when in a state of disagreement with the opponent
(\ref{eqn:condition_for_chi}).

As an example, consider the known extortionate strategy \(p=(8 / 9, 1 / 2, 1 /
3, 0)\) from~\cite{Stewart2012} which is referred to as Extort-2. In
this case, for the standard values of \((R, S, T, P) = (3, 0, 5, 1)\)
constraint (\ref{eqn:condition_for_p1}) corresponds to:

\begin{equation}
    p_1 = \frac{2(p_2 + p_3) + 1}{3}
        = \frac{2(1 / 2 + 1 / 3) + 1}{3}
        = \frac{8}{9}
\end{equation}

It is clear that in this case all constraints hold. As a counterexample,
consider the strategy that cooperates 25\% of the time: \(p=(1 /4, 1 / 4, 1 / 4,
1 / 4)\) obeys~(\ref{eqn:condition_for_chi}) but is not extortionate as:

\begin{equation}
    p_1 \ne \frac{2(p_2 + p_3) + 1}{3}
        = \frac{2(1 / 4 + 1 / 4) + 1}{3}
        = \frac{2}{3}
\end{equation}

\subsection{Measuring Extortion from the History of Play}

Not all strategies are memory-one strategies but it is possible to
measure a given \(p\) from any set of interactions between two strategies.
This approach can then be used to confirm that a given strategy is acting
in an extortionate manner even if it is not a memory-one strategy. However, in
practice, if an exact form for \(p\) is not known but measured from observed
plays of the game then measurement and/or numerical error might lead to an
extortionate strategy not being confirmed as such. \footnote{Comparing theoretic
and actual plays of the IPD is not novel, see for example~\cite{Rand2013}.}

As an example consider Table~\ref{tab:actual_plays_of_ZDextort-2} which shows
some actual plays of Extort-2 (\(p=(8 / 9, 1 / 2, 1 / 3, 0)\)) against an
alternating strategy (\(p=(0, 0, 1, 1)\)). In this particular instance the
measured value of \(p\) for the known extortionate strategy would be:
\((2/2, 1/5, 3/8, 0/4)\) which does not fit the definition of a ZD strategy.

\begin{table}[!hbtp]
    \begin{tabular}{lllllllllllllllllllll}
\toprule
Turn               &  1 &  2 &  3 &  4 &  5 &  6 &  7 &  8 &  9 &  10 &  11 &  12 &  13 &  14 &  15 &  16 &  17 &  18 &  19 &  20 \\\midrule
(8/9, 1/2, 1/3, 0) &  C &  C &  D &  D &  D &  C &  D &  D &  D &   D &   D &   C &   C &   C &   D &   D &   D &   C &   D &   D \\
Alternator         &  C &  D &  C &  D &  C &  D &  C &  D &  C &   D &   C &   D &   C &   D &   C &   D &   C &   D &   C &   D \\
\bottomrule
\end{tabular}

    \caption{A seeded play of 20 turns of two strategies.}
    \label{tab:actual_plays_of_ZDextort-2}
\end{table}

Note that measurement of behaviour might in some cases lead to missing values.
For example the strategy \(p=(8 / 9, 1 / 2, 1 / 3, 0)\) when playing against an
opponent that always cooperates will in fact never visit any state which would allow measurement
\(p_3\) and \(p_4\). To overcome this, it is proposed that if \(s\) is a state
that is not visited then \(p_s\) is approximated using a sensible prior or
imputation. In Section~\ref{sec:numerical-experiments} the overall cooperation
rate is used. Another approach to overcoming this measurement error would be to
measure our strategies in a sufficiently noisy environment.

We can measure how close a strategy is to being zero determinant using standard
linear algebraic approaches. Essentially we attempt to find \(x=(\alpha,
\beta)\) and \(p^*=(\tilde p_1 - 1, \tilde p_2 - 1, \tilde p_3, \tilde p_4)\)
such that

\begin{equation}\label{eqn:linear_algebraic_equation_for_p}
    Cx= p^*
\end{equation}

where \(C\) corresponds to equations
(\ref{eqn:condition_for_tilde_p1}-\ref{eqn:condition_for_tilde_p3}) and is
given by:

\begin{equation}\label{eqn:definition_of_C}
    C =
    \begin{bmatrix}
        R - P & R- P \\
        S - P & T- P \\
        T - P & S- P \\
        0     & 0 \\
    \end{bmatrix}
\end{equation}

Note that in general, equation (\ref{eqn:linear_algebraic_equation_for_p}) will
not necessarily have a solution. From the Rouch\'{e}-Capelli theorem if there is
a solution it is unique since \(\text{rank}(C)=2\) which is the dimension of the
variable \(x\). The best fitting \(x^*\) is defined by:

\begin{equation}\label{eqn:x_star}
    x^* = \text{argmin}_{x\in\mathbb{R}^2}\|C x- p^*\|_2^2
\end{equation}

Known results~\cite{kutner2004applied, rao1973linear, wakefield2013bayesian} yield
$x^*$, corresponding to the nearest extortionate strategy to the
measured \(p\). It is in fact an orthogonal projection of \(p\) on to the plane
defined by (\ref{eqn:condition_for_p1}).

\begin{equation}\label{eqn:x_star_formula}
    x^* = {\left(C^{T}C\right)}^{-1}C^{T}p^{*}
\end{equation}

The squared norm of the remaining error is referred to as sum of squared errors
of prediction (\(\SSe\)):

\begin{equation}\label{eqn:r_squared}
    \SSe = \|C x^*- p^*\|_2^2
\end{equation}

This gives expressions for \(\alpha, \beta\) as \(\alpha=x^*_1\) and
\(\beta=x^*_2\) thus the conditions for a strategy to be acting extortionately
becomes:

\begin{equation}
    -x^*_2 < x^*_1 \label{eqn:measured_condition_for_chi}
\end{equation}

A further known result~~\cite{kutner2004applied, rao1973linear,
wakefield2013bayesian} gives an expression for
\(\SSe\):

\begin{equation}\label{eqn:x_SSError_formula}
    \SSe = {p ^ *} ^ T p ^ * -
           p ^ * C \left(C ^ T C \right) ^ {-1} C ^ T p ^ *
         = {p ^ *} ^ T p ^ * - p ^ * C x ^ *
\end{equation}

Using this approach, the memory-one representation \(p\in\mathbb{R}^4\) of any
strategy against any other can can be measured and if
(\ref{eqn:measured_condition_for_chi}) holds then (\ref{eqn:x_SSError_formula})
can be used to identify if a strategy is acting extortionately. While the
specific memory-one representation might not be one that acts extortionately, a
high \(\SSe\) does imply that a strategy is not extortionate. For a measured
\(p\), \(\SSe\) corresponds to the best fitting \(\alpha, \beta\). Suspicion of
extortion then corresponds to a threshold on \(\SSe\) and a comparison of the
measured \(\chi=\frac{-\beta}{\alpha}\).

\section{Results: Numerical experiments}\label{sec:numerical-experiments}

\cite{Stewart2012} presents results from a tournament with
19strategies
with specific consideration given to ZD strategies. This
tournament is reproduced here using the Axelrod-Python
library~\cite{Knight2016}. To obtain a good measure of the corresponding
transition rates for each strategy all matches have been run for
2000turns and every match has been
repeated 60times. All of this
interaction data is available at~\cite{vincent_knight_2018_1297075}. Note that
in the interest of open scientific practice,~\cite{vincent_knight_2018_1297075}
also contains interaction data for noisy and probabilistic ending interactions
which are not investigated here.

\begin{figure}[!htbp]
    \centering
    \includegraphics[width=.8\textwidth]{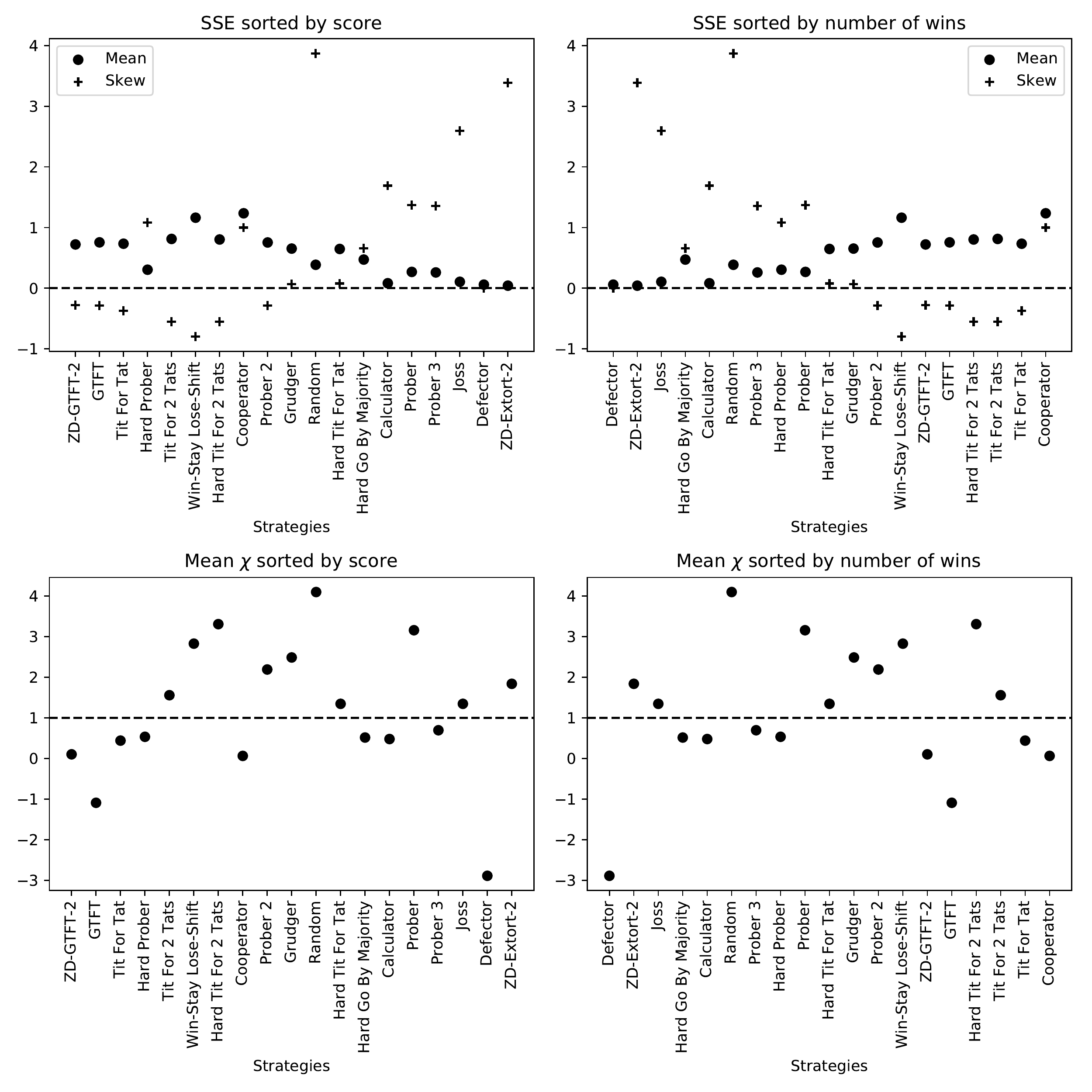}
    \caption{\(\SSe\) and best fitting \(\chi\) for~\cite{Stewart2012},
        ordered both by number of wins and overall score.
        The strategies with a positive skew
        \(\SSe\) and high \(\chi\) win the most matches, although even the known
        extortionate strategy does not act in a perfectly extortionate manner in
        all matches. The strategies with a high score have a negatively skewed
        \(\SSe\).
        }
    \label{fig:sserror_in_stewart_plotkin}
\end{figure}

Figure~\ref{fig:sserror_in_stewart_plotkin} shows the
\(\SSe\) values for all the strategies in the tournament, as
reported in~\cite{Stewart2012} the extortionate strategy Extort-2 gains a large number of
wins. Notice that the mean \(\SSe\) for Extort-2 is approximately zero, while for
the always cooperating strategy Cooperator the \(\SSe\) is far from zero.

\begin{figure}[!htbp]
    \centering
    \includegraphics[width=.95\textwidth]{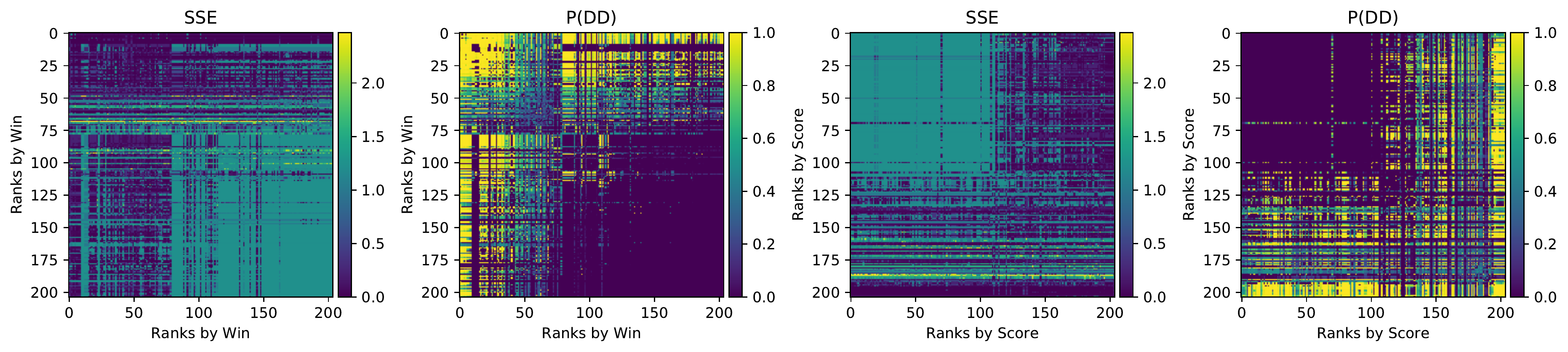}
    \caption{\(\SSe\) and \(P(DD)\) and state probabilities for the strategies for
        the full tournament. The strategies with high number of wins
        have a low \(\SSe\) however are often locked in mutual defection as
        evidenced by a high \(P(DD)\). The strategies with a high score
        have a high \(\SSe\) against the other high scoring strategies
        indicating that fixed linear relationship is being enforced. However
        against the low scoring strategies they have a lower \(\SSe\) and
        against the very lowest scoring strategies a high \(P(DD)\).}
    \label{fig:sse_chi_probabilities_in_full}
\end{figure}

Next we investigate a tournament with
204strategies. The results of
this analysis are shown in Figure~\ref{fig:sse_chi_probabilities_in_full}. The
top ranking strategies by number of wins act in an extortionate way (but not
against all opponents) and it can be seen that a small subgroup of strategies
achieve mutual defection.  All the top ranking strategies according to score
do not extort each other, however they
\textbf{do} exhibit extortionate behaviour towards a number of the lower ranking
strategies.

Note that while a strategy may attempt to act extortionately, not all opponents
can be effectively extorted. For example, a strategy that always defects never
receives a lower score than its opponent. As defined by \cite{Press2012}, an
extortionate ZD strategy will mutually defect with such an opponent which
corresponds to the high values of \(P(DD)\) seen in
Figure~\ref{fig:sse_chi_probabilities_in_full}.

\begin{figure}[!htbp]
    \centering
    \includegraphics[width=\textwidth]{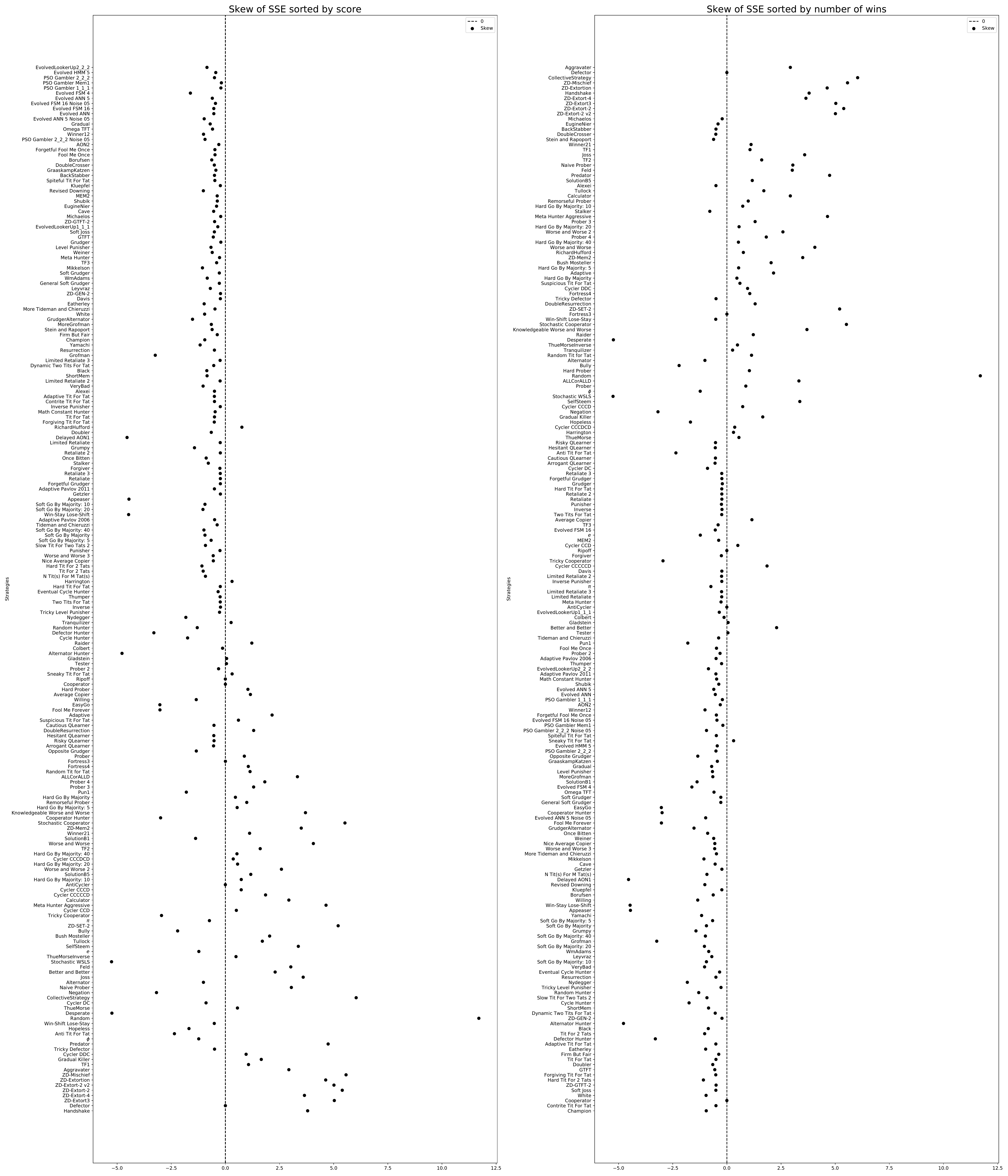}
    \caption{\(\SSe\) for all strategies considered.
        A similar conclusion to that of
        Figure~\ref{fig:sserror_in_stewart_plotkin} can be made: the strategies
        that score highly have a negatively skewed \(\SSe\).}
        \label{fig:sserror_in_std}
\end{figure}

A detailed look at selected strategies is given in
Table~\ref{tbl:overall_summary_results}. The high scoring strategies presented
have a negatively skewed \(\SSe\) whilst the ZD strategies have a low score but
high probability of winning and higher probability of mutual defection.
The skew of \(\SSe\) of all strategies is shown in
Figure~\ref{fig:sserror_in_std} and supports the
same conclusion. This evidences an idea proposed
in~\cite{adami2013evolutionary}: sophisticated strategies are able to recognise
their opponent and defend themselves against extortion.  The high ranking
strategies were in fact trained to maximise score~\cite{Harper2017} which seems
to have created strategies able to extort weaker strategies whilst cooperating
with stronger ones. Indeed unconditional extortion is self defeating.

\begin{table}[!hbtp]
    \begin{center}
    \tiny
    \begin{tabular}{rlrrrrrrr}
\toprule
 Rank &                   Name &  Score per turn &  $P($Win$)$ &  P(DD) &  Median $\chi$ &  Mean SSE &  Skew SSE &  Var SSE \\
\midrule
    1 &   EvolvedLookerUp2\_2\_2 &           2.944 &       0.230 &  0.092 &          0.063 &     1.057 &    -0.857 &    0.160 \\
    2 &          Evolved HMM 5 &           2.944 &       0.205 &  0.110 &          0.063 &     0.796 &    -0.448 &    0.294 \\
    3 &      PSO Gambler 2\_2\_2 &           2.913 &       0.204 &  0.128 &          0.063 &     0.899 &    -0.508 &    0.255 \\
    4 &       PSO Gambler Mem1 &           2.908 &       0.211 &  0.128 &          0.063 &     0.705 &    -0.186 &    0.333 \\
    5 &      PSO Gambler 1\_1\_1 &           2.906 &       0.221 &  0.145 &          0.063 &     0.737 &    -0.209 &    0.296 \\
    7 &          Evolved ANN 5 &           2.893 &       0.225 &  0.185 &          0.063 &     0.804 &    -0.608 &    0.334 \\
   31 &              ZD-GTFT-2 &           2.721 &       0.000 &  0.081 &          0.063 &     0.786 &    -0.502 &    0.289 \\
   45 &               ZD-GEN-2 &           2.689 &       0.016 &  0.096 &          0.063 &     0.694 &    -0.227 &    0.358 \\
   69 &            Tit For Tat &           2.638 &       0.000 &  0.157 &          0.063 &     0.773 &    -0.507 &    0.301 \\
   75 &                 Grumpy &           2.630 &       0.075 &  0.100 &          0.063 &     0.978 &    -1.438 &    0.245 \\
   88 &    Win-Stay Lose-Shift &           2.616 &       0.099 &  0.122 &          0.063 &     1.172 &    -4.501 &    0.027 \\
  103 &  Eventual Cycle Hunter &           2.565 &       0.067 &  0.052 &          0.063 &     0.728 &    -0.338 &    0.357 \\
  127 &               Adaptive &           2.272 &       0.500 &  0.314 &         -1.000 &     0.084 &     2.171 &    0.010 \\
  169 &                  Bully &           1.970 &       0.381 &  0.141 &         -1.000 &     1.373 &    -2.221 &    0.140 \\
  179 &             Alternator &           1.945 &       0.392 &  0.259 &          3.857 &     1.332 &    -1.021 &    0.120 \\
  181 &               Negation &           1.941 &       0.356 &  0.141 &         -1.000 &     1.470 &    -3.204 &    0.083 \\
  182 &     CollectiveStrategy &           1.931 &       0.915 &  0.762 &         -2.888 &     0.085 &     6.082 &    0.028 \\
  183 &              Cycler DC &           1.931 &       0.324 &  0.256 &          3.857 &     1.279 &    -0.900 &    0.140 \\
  188 &               Hopeless &           1.908 &       0.352 &  0.048 &          1.833 &     2.247 &    -1.694 &    0.139 \\
  194 &         Gradual Killer &           1.892 &       0.354 &  0.367 &          0.063 &     0.254 &     1.669 &    0.106 \\
  196 &             Aggravater &           1.879 &       0.930 &  0.739 &         -2.889 &     0.163 &     2.951 &    0.066 \\
  200 &            ZD-Extort-2 &           1.821 &       0.851 &  0.652 &          2.005 &     0.019 &     5.435 &    0.009 \\
  201 &            ZD-Extort-4 &           1.820 &       0.865 &  0.697 &          4.003 &     0.021 &     3.677 &    0.005 \\
  202 &             ZD-Extort3 &           1.810 &       0.862 &  0.687 &          3.028 &     0.015 &     5.066 &    0.005 \\
  203 &               Defector &           1.808 &       0.929 &  0.800 &         -2.889 &     0.059 &     0.000 &    0.000 \\
  204 &              Handshake &           1.806 &       0.870 &  0.737 &         -2.888 &     0.126 &     3.825 &    0.083 \\
\bottomrule
\end{tabular}

    \end{center}
    \caption{Summary of results for a selected list of strategies. Similarly to
        Figure~\ref{fig:sserror_in_stewart_plotkin}, the high scoring strategies
        have a negatively skewed \(\SSe\). The strategies with a
        large number of wins have a low \(\SSe\) and positively skewed
        \(\SSe\). Note that a value of \(\chi=0.063\) and \(\SSe=1.235\)
        corresponds to a vector \(p=(1,1,1,1)\) which highlights that the high
        scoring strategies, adapt and in fact cooperate often.}
    \label{tbl:overall_summary_results}
\end{table}

\section{Evolutionary dynamics}\label{sec:evolutionary-dynamics}

\subsection{Replicator Dynamics}

From the large number of interactions a payoff matrix \(S\)
can be measured where \(S_{ij}\) denotes the score (using standard values of
\((R, S, T, P) = (3, 0, 5, 1)\)) of the \(i\)th strategy against the \(j\)th
strategy. This defines a fitness landscape for which the replicator equation
describes the evolution of a population of strategies:

\begin{equation}\label{eqn:replicator_dynamics}
     \frac{d x_i}{dt} = x_i ((Sx)_i - x^T S x)
\end{equation}

\begin{figure}[!htbp]
    \centering
    \includegraphics[width=.8\textwidth]{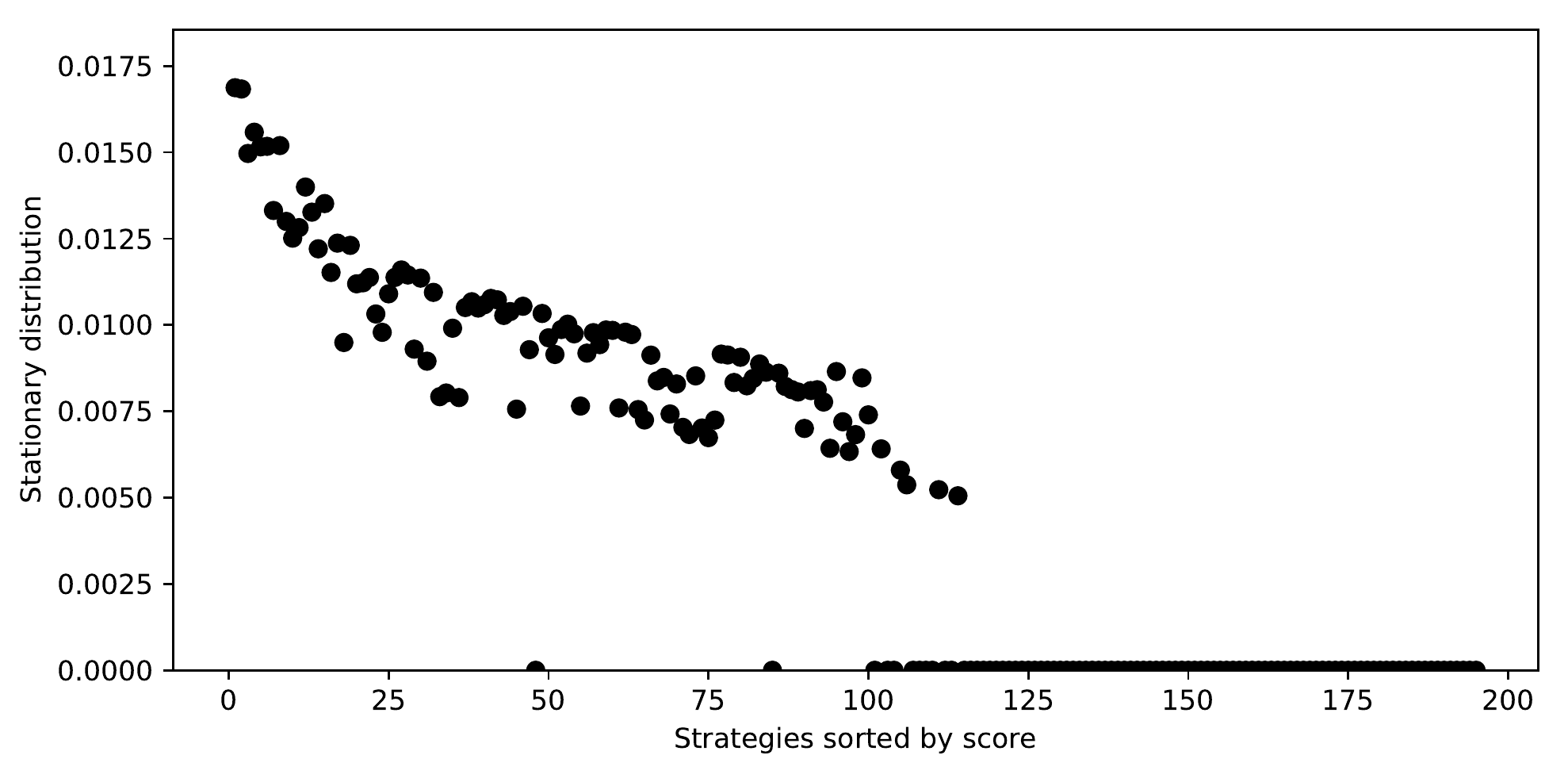}
    \caption{Stationary distribution of the replicator dynamics
    (\ref{eqn:replicator_dynamics}): strategies are ordered by score. Note that
    strategies that make use of the knowledge of the length of the game are
    removed from this analysis as they have an evolutionary advantage.}
    \label{fig:replicator_dynamics}
\end{figure}

Equation (\ref{eqn:replicator_dynamics}) is solved numerically through an
integration technique described in~\cite{Petzold1983} until a stationary vector
\(x=s\) is found.
Figure~\ref{fig:replicator_dynamics} shows the stationary probabilities for each
strategy ranked by score.
It is clear to see that
only the high ranking strategies survive the evolutionary process (in fact,
only 39have a stationary
probability value greater than \(10 ^ {-2}\)).

\begin{figure}[!hbtp]
    \centering
    \includegraphics[width=\textwidth]{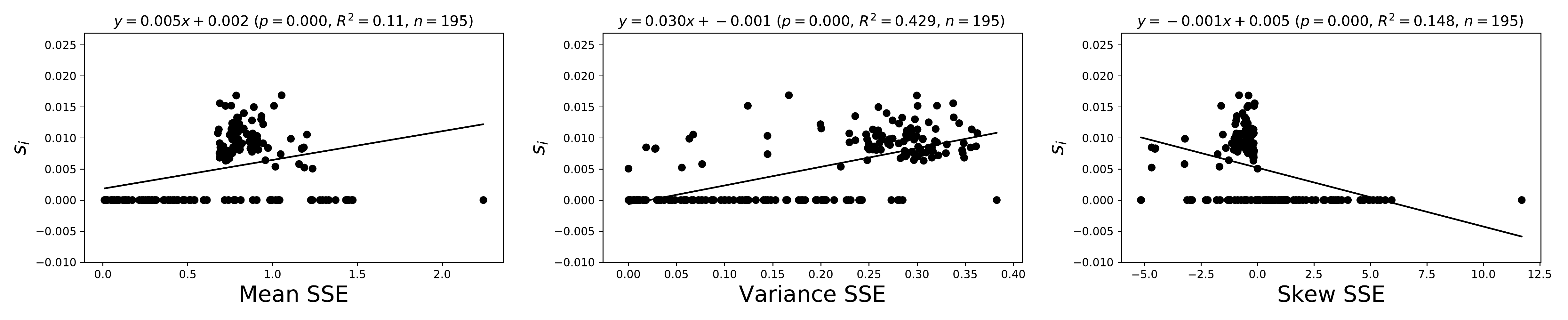}
    \caption{Mean, variance and skew of \(\SSe\) versus the stationary
    probabilities of (\ref{eqn:replicator_dynamics}) a linear regression line is
    included for comparison despite the fact that there is a visible non linear
    relationship. The plot of the skew clearly shows that all high probabilities
    have a negative skew.}
    \label{fig:compare-evolutionary-dynamics-to-sserror}
\end{figure}

Figure~\ref{fig:compare-evolutionary-dynamics-to-sserror} plots the mean and
skew of \(\SSe\) against the stationary probabilities \(s\) of
(\ref{eqn:replicator_dynamics}). Strategies that perform strongly according to
equation (\ref{eqn:replicator_dynamics}) seem to be strategies that have a
negative skew of \(\SSe\): indicating that they often have a high value of
\(\SSe\) (ie do not act extortionately) but have a long left tail allowing them
to adapt when necessary. A general linear model obtained using recursive feature
elimination is shown in Table~\ref{tbl:compare-evolutionary-dynamics-to-sserror}
with stronger predictive power and confirming these conclusions.

\begin{table}[!hbtp]
    \begin{center}
    \tiny
    \begin{center}
\begin{tabular}{lclc}
\toprule
\textbf{Dep. Variable:}    &      $s_i$       & \textbf{  R-squared:         } &    0.648  \\
\textbf{Model:}            &       OLS        & \textbf{  Adj. R-squared:    } &    0.642  \\
\textbf{Method:}           &  Least Squares   & \textbf{  F-statistic:       } &    117.0  \\
\textbf{Date:}             & Fri, 21 Dec 2018 & \textbf{  Prob (F-statistic):} & 5.00e-43  \\
\textbf{Time:}             &     11:01:35     & \textbf{  Log-Likelihood:    } &   851.41  \\
\textbf{No. Observations:} &         195      & \textbf{  AIC:               } &   -1695.  \\
\textbf{Df Residuals:}     &         191      & \textbf{  BIC:               } &   -1682.  \\
\textbf{Df Model:}         &           3      & \textbf{                     } &           \\
\textbf{Covariance Type:}  &    nonrobust     & \textbf{                     } &           \\
\bottomrule
\end{tabular}
\end{center}\begin{center}
\begin{tabular}{lcccccc}
\toprule
                                & \textbf{coef} & \textbf{std err} & \textbf{t} & \textbf{P$>$$|$t$|$} & \textbf{[0.025} & \textbf{0.975]}  \\
\midrule
\textbf{const}                  &       0.0007  &        0.001     &     1.137  &         0.257        &       -0.000    &        0.002     \\
\textbf{('SSE', 'mean')}   &      -0.0134  &        0.002     &    -8.369  &         0.000        &       -0.017    &       -0.010     \\
\textbf{('SSE', 'median')} &       0.0139  &        0.001     &    10.433  &         0.000        &        0.011    &        0.017     \\
\textbf{('SSE', 'var')}    &       0.0069  &        0.003     &     2.402  &         0.017        &        0.001    &        0.013     \\
\bottomrule
\end{tabular}
\end{center}\begin{center}
\begin{tabular}{lclc}
\toprule
\textbf{Omnibus:}       & 17.190 & \textbf{  Durbin-Watson:     } &    1.664  \\
\textbf{Prob(Omnibus):} &  0.000 & \textbf{  Jarque-Bera (JB):  } &   25.453  \\
\textbf{Skew:}          &  0.530 & \textbf{  Prob(JB):          } & 2.97e-06  \\
\textbf{Kurtosis:}      &  4.418 & \textbf{  Cond. No.          } &     23.7  \\
\bottomrule
\end{tabular}
\end{center}
    \end{center}
    \caption{General linear model. This shows that strategies with a low mean
    and high median are more likely to survive the evolutionary dynamics. This
    corresponds to negatively skewed distributions of \(\SSe\) which again
    highlights the importance of adaptability.}
    \label{tbl:compare-evolutionary-dynamics-to-sserror}
\end{table}

\subsection{Finite Population Dynamics: Moran Process}

\begin{figure}[!hbtp]
    \centering
    \includegraphics[width=\textwidth]{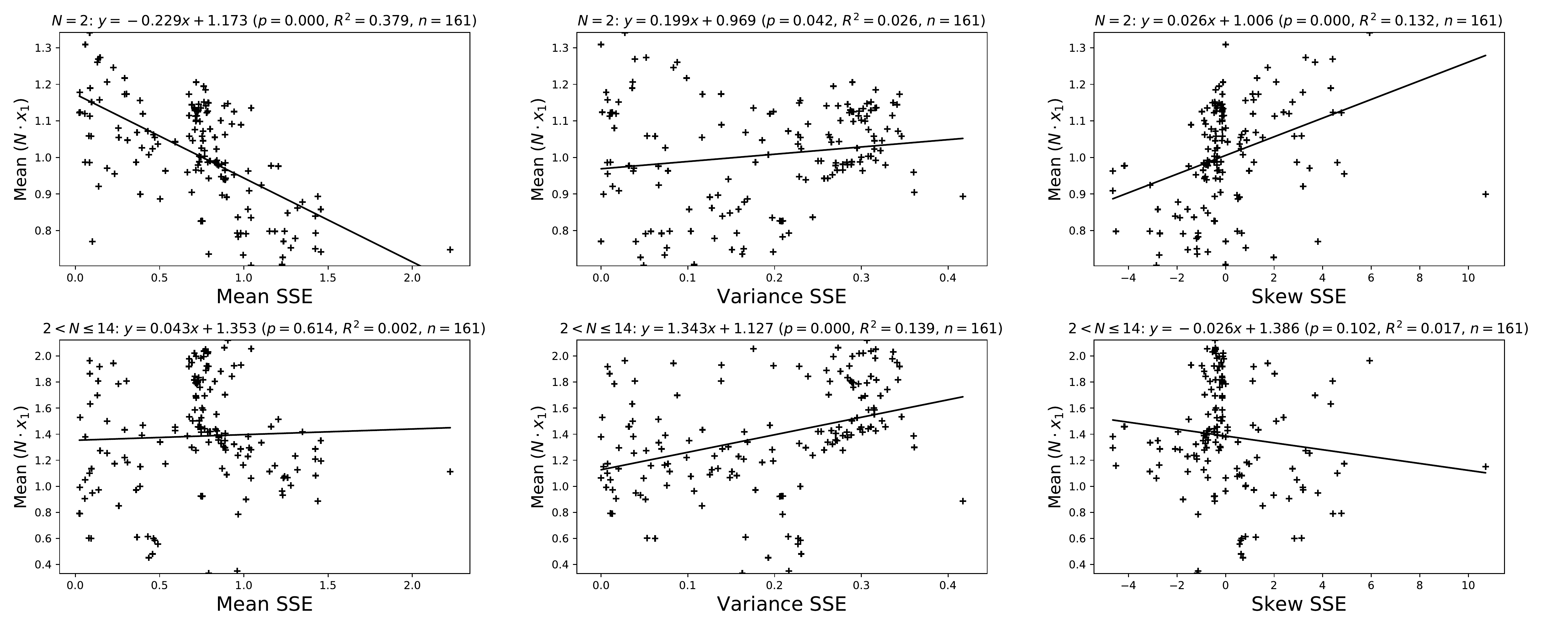}
    \caption{The mean, variance and skew of
    \(\SSe\) against the normalised pairwise fixation probabilities
    from~\cite{Moran1707} (for a given strategy averaged over all opponents and
    population sizes). As for
    Figure~\ref{fig:compare-evolutionary-dynamics-to-sserror} the linear
    regression lines are include for comparison despite there being no clear
    linear relationship. The clustering either side of a value of skew equal to
    0 show that strategies with above neutral
    fixation (\(N\cdot x_1>1\)) negative skew.}
    \label{fig:compare-fixation-to-sserror}
\end{figure}

In~\cite{Moran1707} a large data set of pairwise fixation probabilities in the
Moran process is made available at~\cite{vincent_knight_2017_1040129}
Figure~\ref{fig:compare-fixation-to-sserror} shows linear models fitted to three
summary measures of \(\SSe\) and the mean (over population size \(N\) and
opponents) value of \(x_1\cdot N\). This specific measure of fixation is chosen
as \(x_1\) is usually compared to the neutral fixation probability of \(1 / N\).
As was noted in~\cite{Moran1707}, the specific case of \(N=2\) differs from all
other population sizes which is why it is presented in isolation. Similarly to
the conclusions from Figure~\ref{fig:compare-evolutionary-dynamics-to-sserror}
we note that there is a significant relationship between the skew of
\(\SSe\) and the ability for a strategy to become fixed.
A general linear model obtained through recursive feature elimination is shown
in Table~\ref{tbl:compare-fixation-to-sserror} which confirms the conclusions.

\begin{table}[!hbtp]
    \begin{center}
    \tiny
    \begin{center}
\begin{tabular}{lclc}
\toprule
\textbf{Dep. Variable:}    &       mean       & \textbf{  R-squared:         } &    0.319  \\
\textbf{Model:}            &       OLS        & \textbf{  Adj. R-squared:    } &    0.310  \\
\textbf{Method:}           &  Least Squares   & \textbf{  F-statistic:       } &    36.53  \\
\textbf{Date:}             & Fri, 21 Dec 2018 & \textbf{  Prob (F-statistic):} & 9.74e-14  \\
\textbf{Time:}             &     10:42:28     & \textbf{  Log-Likelihood:    } &  -42.272  \\
\textbf{No. Observations:} &         159      & \textbf{  AIC:               } &    90.54  \\
\textbf{Df Residuals:}     &         156      & \textbf{  BIC:               } &    99.75  \\
\textbf{Df Model:}         &           2      & \textbf{                     } &           \\
\textbf{Covariance Type:}  &    nonrobust     & \textbf{                     } &           \\
\bottomrule
\end{tabular}
\end{center}\begin{center}
\begin{tabular}{lcccccc}
\toprule
                                & \textbf{coef} & \textbf{std err} & \textbf{t} & \textbf{P$>$$|$t$|$} & \textbf{[0.025} & \textbf{0.975]}  \\
\midrule
\textbf{const}                  &       1.2815  &        0.056     &    22.993  &         0.000        &        1.171    &        1.392     \\
\textbf{('SSE', 'mean')}   &      -1.0620  &        0.145     &    -7.323  &         0.000        &       -1.348    &       -0.776     \\
\textbf{('SSE', 'median')} &       0.9037  &        0.106     &     8.535  &         0.000        &        0.695    &        1.113     \\
\bottomrule
\end{tabular}
\end{center}\begin{center}
\begin{tabular}{lclc}
\toprule
\textbf{Omnibus:}       &  2.302 & \textbf{  Durbin-Watson:     } &    1.716  \\
\textbf{Prob(Omnibus):} &  0.316 & \textbf{  Jarque-Bera (JB):  } &    1.850  \\
\textbf{Skew:}          & -0.199 & \textbf{  Prob(JB):          } &    0.397  \\
\textbf{Kurtosis:}      &  3.348 & \textbf{  Cond. No.          } &     11.2  \\
\bottomrule
\end{tabular}
\end{center}
    \end{center}
    \caption{General linear model. This shows that strategies with a high mean
        and low median are likely to be evolutionarily stable. This corresponds
        to negatively skewed distributions of \(\SSe\) which again highlights
        the importance of adaptability.}
    \label{tbl:compare-fixation-to-sserror}
\end{table}

These findings confirm the work of~\cite{Moran1707} in which sophisticated
strategies resist evolutionary invasion of shorter memory strategies. This also
confirms the work of~\cite{adami2013evolutionary, hilbe2015partners} which
proved that ZD strategies where not evolutionarily stable due to the fact that
they score poorly against themselves.

The work also provides strong evidence to the importance of adaptability:
strategies that offer a variety of behaviours corresponding to a higher standard
deviation of \(\SSe\) are significantly more likely to survive the
evolutionary process. This corresponds to the following quote
of~\cite{darwin1869origin}:

\begin{quote}
``It is not the most intellectual of the species that survives; it is not the
strongest that survives; but the species that survives is the one that is able
to adapt to and to adjust best to the changing environment in which it finds
itself.''
\end{quote}

\section{Discussion}\label{sec:conclusion}

This work defines an approach to measure whether or not a player is using an
extortionate strategy as defined in~\cite{Press2012}, or a strategy that behaves
similarly, broadening the definition of extortionate behavior. All extortionate
strategies have been classified as lying on a triangular plane. This rigorous
classification fails to be robust to small measurement error, thus a statistical
approach is proposed approximating the solution of a linear system. 
This method
was applied to a large number of pairwise interactions.

The work of~\cite{Press2012}, while showing that a clever approach to taking
advantage of another memory-one strategy exists, is not the full story.
Though the elegance of this result is very attractive, just as the simplicity of
the victory of Tit For Tat in Axelrod's original tournaments was, it is
incomplete and in the author's opinions, has been oversimplified and
overgeneralized in subsequent work. Extortionate strategies achieve a high
number of wins but they do generally not achieve a high score and fail to be
evolutionarily stable.

Rather more sophisticated strategies are able to adapt to a variety of opponents
and act extortionately only against weaker strategies while cooperating with
like-minded strategies that are not susceptible to extortion. This adaptability
may be key to maintaining sustained cooperation, as some of these strategies
emerged naturally from evolutionary processes trained to maximize payoff in
IPD tournaments and fixation in population dynamics.

Following Axelrod's seminal work~\cite{Axelrod1980, Axelrod1980a}, it was
commonly thought that evolutionary cooperation required strategies that followed
a simple set of rules. The discovery/definition of extortionate
strategies~\cite{Press2012} seemingly showed that complex strategies could be
taken advantage of. In this manuscript it has been shown that not only is it
possible to detect and prevent extortionate behaviour but that more complex
strategies can be evolutionary stable. The complex strategies in question were
obtained through reinforcement learning approaches~\cite{Harper2017, Moran1707}.
Thus, this demonstrates that it is possible to recognise extortion, both
theoretically using \(\SSe\) but also that this ability can develop through
reinforcement learning. It seems human difficulty in directly developing
effective complex strategies has been incorrectly generalized to a weakness
in complex strategies themselves, which is demonstrable not the case. In fact,
complex strategies can be the most effective against a diverse set of opponents.

In closing, the authors wish to emphasize the role of comprehensive simulations to temper
theoretical results from overgeneralization, and perhaps more importantly, the
ability of simulations to provide insights that are difficult to obtain from theory.

\section*{Acknowledgements}

The following open source software libraries were used in this research:

\begin{itemize}
    \item The Axelrod ~\cite{Knight2016, Knight2018} library (IPD strategies and
        tournaments).
    \item The sympy library~\cite{Meurer2017} (verification of all symbolic
        calculations).
    \item The matplotlib~\cite{Droettboom2018} library (visualisation).
    \item The pandas~\cite{Structures2010}, dask~\cite{Dask2016} and
        NumPy~\cite{Oliphant2015} libraries (data manipulation).
    \item The SciPy~\cite{Jones2001} library (numerical integration of the
        replicator equation).
\end{itemize}

This work was performed using the computational facilities of the Advanced
Research Computing @ Cardiff (ARCCA) Division, Cardiff University.

\section*{Author contributions}

VK and NG conceived the idea. MH, JG, NG and VK were all involved in carrying
out the research and writing the manuscript.

\printbibliography



\section*{Supplementary material}

# Proof of algebraic condition for extortionate strategies

The defining equations for an extortionate strategy are:

$$\tilde{p}_1 = \alpha(R - P) + \beta(R - P) \tag{1}$$
$$\tilde{p}_2 = \alpha(S - P) + \beta(T - P) \tag{2}$$
$$\tilde{p}_3 = \alpha(T - P) + \beta(S - P) \tag{3}$$
$$\tilde{p}_4 = 0 \tag{4}$$

Using equation (2), $\alpha$ is isolated

$$\alpha = \frac{-\beta(P - T) - \tilde{p}_2}{P - S} \tag{5}$$

Substituting this value in to equation (3), $\beta$ is isolated:

$$\beta = -\frac{P\tilde{p}_1 - P\tilde{p}_2 + S\tilde{p}_2 - T\tilde{p}_1}{(S - T)(2P - S - T)} \tag{6}$$

Substituting this back in to (5) gives:

$$\alpha = \frac{-\tilde{p}_2 + (P - T)(P\tilde{p}_1 - P\tilde{p}_2 + S\tilde{p}_2 - T\tilde{p}_1)}{(S - T)(2P - S - T)(P - S)} \tag{7}$$

Substituting equations (6-7) in to equation (1) gives the required expression for $p_1$.
Taking the ratio of equations (6-7) gives the required expression for $\chi$.
Finally, the condition $\chi > 1$ corresponds to:

$$\tilde{p}_2(P - T) + \tilde{p}_3(S - P) > \tilde{p}_2(P - S) + \tilde{p}_3(T - P) \tag{8}$$

which can be simplified to:

$$\tilde{p}_2 < -\tilde{p}_3 \tag{9}$$

recalling that $\tilde{p}_2 = p_2 - 1$ and $\tilde{p}_3 = p_3$ gives the required result.

# List of all strategies used from [21]

1. Adaptive - Deterministic - Memory length: ∞ - [25]
2. Adaptive Tit For Tat: 0.5 - Deterministic - Memory length: ∞ - [39]
3. Aggravater - Deterministic - Memory length: ∞ - [21]
4. Alexei: (D,) - Deterministic - Memory length: ∞ - [43]
5. ALLCorALLD - Stochastic - Memory length: 1 - [2]
6. Alternator - Deterministic - Memory length: 1 - [11, 31]
7. Alternator Hunter - Deterministic - Memory length: ∞ - [21]
8. AntiCycler - Deterministic - Memory length: ∞ - [21]
9. Anti Tit For Tat - Deterministic - Memory length: 1 - [18]
10. AON2 - Deterministic - Memory length: 2 - [19]
11. Adaptive Pavlov 2006 - Deterministic - Memory length: ∞ - [24]
12. Adaptive Pavlov 2011 - Deterministic - Memory length: ∞ - [25]
13. Appeaser - Deterministic - Memory length: ∞ - [21]
14. Arrogant QLearner - Stochastic - Memory length: ∞ - [21]
15. Average Copier - Stochastic - Memory length: ∞ - [21]
16. BackStabber: (D, D) - Deterministic - Memory length: ∞ - [21]
17. Better and Better - Stochastic - Memory length: ∞ - [29]
18. Black - Stochastic - Memory length: 5 - [10]
19. Borufsen - Deterministic - Memory length: ∞ - [10]
20. Bully - Deterministic - Memory length: 1 - [32]
21. Bush Mosteller: 0.5, 0.5, 3.0, 0.5 - Stochastic - Memory length: ∞ - [20]
22. Calculator - Stochastic - Memory length: ∞ - [29]
23. Cautious QLearner - Stochastic - Memory length: ∞ - [21]
24. Cave - Stochastic - Memory length: ∞ - [10]
25. Champion - Stochastic - Memory length: ∞ - [10]
26. Colbert - Deterministic - Memory length: 4 - [10]
27. CollectiveStrategy - Deterministic - Memory length: ∞ - [26]
28. Contrite Tit For Tat - Deterministic - Memory length: 3 - [42]
29. Cooperator - Deterministic - Memory length: 0 - [11, 31, 34]
30. Cooperator Hunter - Deterministic - Memory length: ∞ - [21]
31. Cycle Hunter - Deterministic - Memory length: ∞ - [21]
32. Cycler CCCCCD - Deterministic - Memory length: 5 - [21]
33. Cycler CCCD - Deterministic - Memory length: 3 - [21]
34. Cycler CCD - Deterministic - Memory length: 2 - [31]
35. Cycler DC - Deterministic - Memory length: 1 - [21]
36. Cycler DDC - Deterministic - Memory length: 2 - [31]
37. Cycler CCCDCD - Deterministic - Memory length: 5 - [21]
38. Davis: 10 - Deterministic - Memory length: ∞ - [9]
39. Defector - Deterministic - Memory length: 0 - [11, 31, 34]
40. Defector Hunter - Deterministic - Memory length: ∞ - [21]
41. Desperate - Stochastic - Memory length: 1 - [41]
42. Delayed AON1 - Deterministic - Memory length: 2 - [19]
43. DoubleCrosser: (D, D) - Deterministic - Memory length: ∞ - [21]
44. Doubler - Deterministic - Memory length: ∞ - [29]
45. DoubleResurrection - Deterministic - Memory length: 5 - [15]
46. EasyGo - Deterministic - Memory length: ∞ - [29, 25]
47. Eatherley - Stochastic - Memory length: ∞ - [10]
48. EugineNier: (D,) - Deterministic - Memory length: ∞ - [43]
49. Eventual Cycle Hunter - Deterministic - Memory length: ∞ - [21]
50. Evolved ANN - Deterministic - Memory length: ∞ - [21]
51. Evolved ANN 5 - Deterministic - Memory length: ∞ - [21]
52. Evolved ANN 5 Noise 05 - Deterministic - Memory length: ∞ - [21]
53. Evolved FSM 4 - Deterministic - Memory length: ∞ - [21]
54. Evolved FSM 16 - Deterministic - Memory length: ∞ - [21]
55. Evolved FSM 16 Noise 05 - Deterministic - Memory length: ∞ - [21]
56. EvolvedLookerUp1_1_1 - Deterministic - Memory length: ∞ - [21]
57. EvolvedLookerUp2_2_2 - Deterministic - Memory length: ∞ - [21]
58. Evolved HMM 5 - Stochastic - Memory length: 5 - [21]
59. Feld: 1.0, 0.5, 200 - Stochastic - Memory length: 200 - [9]
60. Firm But Fair - Stochastic - Memory length: 1 - [16]
61. Fool Me Forever - Deterministic - Memory length: ∞ - [21]
62. Fool Me Once - Deterministic - Memory length: ∞ - [21]
63. Forgetful Fool Me Once: 0.05 - Stochastic - Memory length: ∞ - [21]
64. Forgetful Grudger - Deterministic - Memory length: 10 - [21]
65. Forgiver - Deterministic - Memory length: ∞ - [21]
66. Forgiving Tit For Tat - Deterministic - Memory length: ∞ - [21]
67. Fortress3 - Deterministic - Memory length: 2 - [7]
68. Fortress4 - Deterministic - Memory length: 3 - [7]
69. GTFT: 0.33 - Stochastic - Memory length: 1 - [33, 17]
70. General Soft Grudger: n=1,d=4,c=2 - Deterministic - Memory length: ∞ - [21]
71. Getzler - Stochastic - Memory length: ∞ - [10]
72. Gladstein - Deterministic - Memory length: ∞ - [10]
73. Soft Go By Majority - Deterministic - Memory length: ∞ - [11, 31, 10]
74. Soft Go By Majority: 10 - Deterministic - Memory length: 10 - [21]
75. Soft Go By Majority: 20 - Deterministic - Memory length: 20 - [21]
76. Soft Go By Majority: 40 - Deterministic - Memory length: 40 - [21]

77. Soft Go By Majority: 5 - Deterministic - Memory length: 5 - [21]
78. φ - Deterministic - Memory length: ∞ - [21]
79. GraaskampKatzen - Deterministic - Memory length: ∞ - [10]
80. Gradual - Deterministic - Memory length: ∞ - [13]
81. Gradual Killer: (D, D, D, D, D, C, C) - Deterministic - Memory length: ∞ - [29]
82. Grofman - Stochastic - Memory length: ∞ - [9]
83. Grudger - Deterministic - Memory length: 1 - [25, 9, 41, 12, 13]
84. GrudgerAlternator - Deterministic - Memory length: ∞ - [29]
85. Grumpy: Nice, 10, -10 - Deterministic - Memory length: ∞ - [21]
86. Handshake - Deterministic - Memory length: ∞ - [35]
87. Hard Go By Majority - Deterministic - Memory length: ∞ - [31]
88. Hard Go By Majority: 10 - Deterministic - Memory length: 10 - [21]
89. Hard Go By Majority: 20 - Deterministic - Memory length: 20 - [21]
90. Hard Go By Majority: 40 - Deterministic - Memory length: 40 - [21]
91. Hard Go By Majority: 5 - Deterministic - Memory length: 5 - [21]
92. Hard Prober - Deterministic - Memory length: ∞ - [29]
93. Hard Tit For 2 Tats - Deterministic - Memory length: 3 - [38]
94. Hard Tit For Tat - Deterministic - Memory length: 3 - [40]
95. Harrington - Stochastic - Memory length: ∞ - [10]
96. Hesitant QLearner - Stochastic - Memory length: ∞ - [21]
97. Hopeless - Stochastic - Memory length: 1 - [41]
98. Inverse - Stochastic - Memory length: ∞ - [21]
99. Inverse Punisher - Deterministic - Memory length: ∞ - [21]
100. Joss: 0.9 - Stochastic - Memory length: 1 - [38, 9]
101. Kluepfel - Stochastic - Memory length: ∞ - [10]
102. Knowledgeable Worse and Worse - Stochastic - Memory length: ∞ - [21]
103. Level Punisher - Deterministic - Memory length: ∞ - [15]
104. Leyvraz - Stochastic - Memory length: 3 - [10]
105. Limited Retaliate: 0.1, 20 - Deterministic - Memory length: ∞ - [21]
106. Limited Retaliate 2: 0.08, 15 - Deterministic - Memory length: ∞ - [21]
107. Limited Retaliate 3: 0.05, 20 - Deterministic - Memory length: ∞ - [21]
108. Math Constant Hunter - Deterministic - Memory length: ∞ - [21]
109. Naive Prober: 0.1 - Stochastic - Memory length: 1 - [25]
110. MEM2 - Deterministic - Memory length: ∞ - [27]
111. Michaelos: (D,) - Stochastic - Memory length: ∞ - [43]
112. Mikkelson - Deterministic - Memory length: ∞ - [10]
113. MoreGrofman - Deterministic - Memory length: 8 - [10]
114. More Tideman and Chieruzzi - Deterministic - Memory length: ∞ - [10]
115. Negation - Stochastic - Memory length: 1 - [40]
116. Nice Average Copier - Stochastic - Memory length: ∞ - [21]
117. N Tit(s) For M Tat(s): 3, 2 - Deterministic - Memory length: 3 - [21]
118. Nydegger - Deterministic - Memory length: 3 - [9]
119. Omega TFT: 3, 8 - Deterministic - Memory length: ∞ - [37]
120. Once Bitten - Deterministic - Memory length: 12 - [21]
121. Opposite Grudger - Deterministic - Memory length: ∞ - [21]
122. π - Deterministic - Memory length: ∞ - [21]
123. Predator - Deterministic - Memory length: ∞ - [7]
124. Prober - Deterministic - Memory length: ∞ - [25]
125. Prober 2 - Deterministic - Memory length: ∞ - [29]
126. Prober 3 - Deterministic - Memory length: ∞ - [29]
127. Prober 4 - Deterministic - Memory length: ∞ - [29]
128. Pun1 - Deterministic - Memory length: ∞ - [6]
129. PSO Gambler 1_1_1 - Stochastic - Memory length: ∞ - [21]
130. PSO Gambler 2_2_2 - Stochastic - Memory length: ∞ - [21]
131. PSO Gambler 2_2_2 Noise 05 - Stochastic - Memory length: ∞ - [21]
132. PSO Gambler Mem1 - Stochastic - Memory length: 1 - [21]
133. Punisher - Deterministic - Memory length: ∞ - [21]
134. Raider - Deterministic - Memory length: ∞ - [8]
135. Random: 0.5 - Stochastic - Memory length: 0 - [39, 9]
136. Random Hunter - Deterministic - Memory length: ∞ - [21]
137. Random Tit for Tat: 0.5 - Stochastic - Memory length: 1 - [21]
138. Remorseful Prober: 0.1 - Stochastic - Memory length: 2 - [25]
139. Resurrection - Deterministic - Memory length: 5 - [15]
140. Retaliate: 0.1 - Deterministic - Memory length: ∞ - [21]
141. Retaliate 2: 0.08 - Deterministic - Memory length: ∞ - [21]
142. Retaliate 3: 0.05 - Deterministic - Memory length: ∞ - [21]
143. Revised Downing: True - Deterministic - Memory length: ∞ - [9]
144. RichardHufford - Deterministic - Memory length: ∞ - [10]
145. Ripoff - Deterministic - Memory length: 3 - [5]
146. Risky QLearner - Stochastic - Memory length: ∞ - [21]
147. SelfSteem - Stochastic - Memory length: ∞ - [14]
148. ShortMem - Deterministic - Memory length: 10 - [14]
149. Shubik - Deterministic - Memory length: ∞ - [9]
150. Slow Tit For Two Tats 2 - Deterministic - Memory length: 2 - [29]
151. Sneaky Tit For Tat - Deterministic - Memory length: ∞ - [21]
152. Soft Grudger - Deterministic - Memory length: 6 - [25]
153. Soft Joss: 0.9 - Stochastic - Memory length: 1 - [29]
154. SolutionB1 - Deterministic - Memory length: 2 - [4]
155. SolutionB5 - Deterministic - Memory length: ∞ - [4]
156. Spiteful Tit For Tat - Deterministic - Memory length: ∞ - [29]
157. Stalker: (D,) - Stochastic - Memory length: ∞ - [14]
158. Stein and Rapoport: 0.05: (D, D) - Deterministic - Memory length: ∞ - [9]
159. Stochastic Cooperator - Stochastic - Memory length: 1 - [1]
160. Stochastic WSLS: 0.05 - Stochastic - Memory length: 1 - [3]
161. Suspicious Tit For Tat - Deterministic - Memory length: 1 - [13, 18]
162. Tester - Deterministic - Memory length: ∞ - [10]

163. TF1 - Deterministic - Memory length: ∞ - [21]
164. TF2 - Deterministic - Memory length: ∞ - [21]
165. TF3 - Deterministic - Memory length: ∞ - [21]
166. ThueMorse - Deterministic - Memory length: ∞ - [21]
167. ThueMorseInverse - Deterministic - Memory length: ∞ - [21]
168. Thumper - Deterministic - Memory length: ∞ - [5]
169. Tideman and Chieruzzi - Deterministic - Memory length: ∞ - [9]
170. Tit For Tat - Deterministic - Memory length: 1 - [9]
171. Tit For 2 Tats - Deterministic - Memory length: 2 - [11]
172. Tranquilizer - Stochastic - Memory length: ∞ - [9]
173. Tricky Cooperator - Deterministic - Memory length: 10 - [21]
174. Tricky Defector - Deterministic - Memory length: ∞ - [21]
175. Tricky Level Punisher - Deterministic - Memory length: ∞ - [15]
176. Tullock: 11 - Stochastic - Memory length: 11 - [9]
177. Two Tits For Tat - Deterministic - Memory length: 2 - [11]
178. VeryBad - Deterministic - Memory length: ∞ - [14]
179. Weiner - Deterministic - Memory length: ∞ - [10]
180. White - Deterministic - Memory length: ∞ - [10]
181. Willing - Stochastic - Memory length: 1 - [41]
182. Winner12 - Deterministic - Memory length: 2 - [30]
183. Winner21 - Deterministic - Memory length: 2 - [30]
184. Win-Shift Lose-Stay: D - Deterministic - Memory length: 1 - [25]
185. Win-Stay Lose-Shift: C - Deterministic - Memory length: 1 - [33, 38, 22]
186. WmAdams - Stochastic - Memory length: ∞ - [10]
187. Worse and Worse - Stochastic - Memory length: ∞ - [29]
188. Worse and Worse 2 - Stochastic - Memory length: ∞ - [29]
189. Worse and Worse 3 - Stochastic - Memory length: ∞ - [29]
190. Yamachi - Deterministic - Memory length: ∞ - [10]
191. ZD-Extortion: 0.2, 0.1, 1 - Stochastic - Memory length: 1 - [36]
192. ZD-Extort-2: 0.11111111111111111, 0.5 - Stochastic - Memory length: 1 - [38]
193. ZD-Extort3: 0.11538461538461539, 0.3333333333333333, 1 - Stochastic - Memory length: 1 - [34]
194. ZD-Extort-2 v2: 0.125, 0.5, 1 - Stochastic - Memory length: 1 - [23]
195. ZD-Extort-4: 0.23529411764705882, 0.25, 1 - Stochastic - Memory length: 1 - [21]
196. ZD-GTFT-2: 0.25, 0.5 - Stochastic - Memory length: 1 - [38]
197. ZD-GEN-2: 0.125, 0.5, 3 - Stochastic - Memory length: 1 - [23]
198. ZD-Mem2 - Stochastic - Memory length: 2 - [28]
199. ZD-Mischief: 0.1, 0.0, 1 - Stochastic - Memory length: 1 - [36]
200. ZD-SET-2: 0.25, 0.0, 2 - Stochastic - Memory length: 1 - [23]
201. e - Deterministic - Memory length: ∞ - [21]
202. Dynamic Two Tits For Tat - Stochastic - Memory length: ∞ - [21]
203. Meta Hunter: 6 players - Deterministic - Memory length: ∞ - [21]
204. Meta Hunter Aggressive: 7 players - Deterministic - Memory length: ∞ - [21]



\end{document}